\documentstyle[12pt]{article}

\evensidemargin=\oddsidemargin
\begin{document}
\begin{flushright}
{\large \sf  VPI-IHEP-93-12\\
             hep-ph/9402293\\}
\end{flushright}
\bigskip
\begin{center}
{\large \bf Inconsistency of QED in the Presence of Dirac Monopoles
\footnote{Work supported in part by the U.S. Department of
           Energy under grant DEFG05-92ER40709.} }\\[1.5cm]

{\bf Hong-Jian He }$^{a,b}$ ~~~~{\bf Zhaoming Qiu }$^b$
{}~~~~{\bf Chia-Hsiung Tze }$^a$  \\[0.7cm]

$^a$ Department of Physics and Institute for High Energy Physics\\
     Virginia Polytechnic Institute and State University\\
     Blacksburg, Virginia 24061-0435, U.S.A.
 \footnote{Present address of H.J. He ( E-mail: hjhe@vtinte.phys.vt.edu ).}\\

$^b$ China Center of Advanced Science and Technology (World Laboratory) \\
     P. O. Box 8730, Beijing 100080, P.R. China
\end{center}

\bigskip
\centerline{\bf Abstract}
\bigskip
\begin{center}
\begin{minipage}[c]{15cm}
\begin{sf}
\parindent=1.2cm

\noindent
A precise formulation of $U(1)$ local gauge invariance in QED is presented,
which clearly shows that the gauge coupling associated with
the unphysical longitudinal photon field is non-observable and actually
has an arbitrary value. We then re-examine the Dirac quantization
condition  and find that its derivation involves solely the unphysical
longitudinal coupling. Hence an inconsistency inevitably arises
in the presence of Dirac monopoles and this can be considered as
a theoretical evidence  against their  existence.
An alternative, independent proof of this conclusion is also presented.

\end{sf}
\end{minipage}
\end{center}
\vspace{1.0cm}
\begin{center}
(~~ Published in Zeitschrift f$\ddot{u}$r Physik {\bf C65}, p.175-182
               ~~)
\end{center}

\newpage
\begin{sf}
\noindent
{\bf 1. Introduction }
\vspace{0.2cm}

The concept of magnetic monopoles dates back to the dawn of
the science of magnetism. The existence of monopoles seems to be a
natural generalization of the classical Maxwell theory (1864) on the
ground of dual symmetry.
Sixty-seven  years later, Dirac pioneered
a theory of point monopoles$^{[1]}$. His formulation,
which puts monopoles by hand in QED, requires an Abelian gauge potential
with string singularities and results in the
famous Dirac quantization condition (DQC), accounting notably for
electric charge quantization. Dirac's work has certainly been a main
driving force behind numerious unsuccessful experimental searches
for monopoles$^{[2]}$. This state of affairs has led us to ask whether
there may still be some deep, unexplored theoretical
reason(s) behind these {\it negative} experimental results.

Over the years, in spite of the continued non-observation
of Dirac monopoles, the subtleties and proper handlings of Dirac string
singularities in quantized
theory have been extensively studied to confirm the consistency
of Dirac monopoles with quantum physics. The immediate implications of
the DQC are two-fold: (1). electric charges are quantized even if one
monopole exists; (2). magnetic monopoles interact strongly and therefore
from energetic consideration should be quantum mechanically extended objects.
Yet, in ordinary Maxwell theory, the condition
$~~\nabla\cdot (\nabla\times\vec{A})=0~~$, while allowing,
via albeit a {\it singular} potential, point-like monopoles, realized as
endpoints of Dirac string singularities, precludes a spread-out magnetic
density. There lies in our view the weak point of the Abelian theory.
As we now know, the only way such a potential can be non-singular is for it
to be embedded in a non-Abelian type gauge theory ( although at present a
realistic unification theory of such kind is still absent ).
Indeed t' Hooft-Polyakov monopoles$^{[3]}$
obeying the DQC arise naturally as finite energy,
spatially extended solutions of broken non-Abelian gauge theories.
In fact, it is precisely this striking consistency between the extension
and the DQC of the solitonic monopoles in the latter theories that again
motivates us to carefully re-examine the consistency question of Dirac's
theory.

In this paper we analyze the inconsistency of QED with Dirac
point monopoles.  We found that
the presence of Dirac monopoles will force the non-observable gauge
coupling associated with the unphysical longitudinal gauge degree of
freedom of
the photon field to be exactly the same as that associated with the two
physical transverse degrees of freedom.
This unfortunately violates the general spirit of the local gauge invariance
and implies that the longitudinal photon may have some physical consequences.
We see this as a clear evidence that the unphysical Dirac strings cannot
be  really non-observable.
   Furthermore, an alternative, independent proof of the inconsistecy is
presented.
We enlarge the local gauge invariance of QED from $~U(1)_{em}~$ to
$~U(1)_{em} \times U(1)_{\Theta}~$ by introducing another unphysical
pure gauge field $~\Theta~$ with an independent, unphysical gauge coupling
$~\tilde{e}~$. This pure gauge field can be gauge-transformed away and
the resulting theory is proved to be {\it identical to} standard QED.
We then re-examine the Dirac quantization condition (DQC) for point
monopoles and find that two essentially different DQCs can be derived.
One DQC involves a
gauge coupling $~e~$ in the $~U(1)_{em}~$ group and the other
only the unphysical gauge coupling $~\tilde{e}~$ in the
$~U(1)_{\Theta}~$ group. The unique physically consistent solution of these
two DQCs is a vanishing magnetic charge, which implies that no Dirac
monopole exists in nature.
A precise formulation of $~U(1)_{em}~$
local gauge invariance is presented in
Sec.2 and the DQC is carefully re-examined in Sec.3.
The alternative approach is analyzed in Sec.4.
Other aspects of such
inconsistencies of Dirac monopoles have also been found. They are examined
elsewhere$^{[4][5]}$.

\vspace{0.5cm}
\noindent
{\bf 2. The precise formulation of local $U(1)_{em}$ gauge invariance
        and the  Ward-Takahashi identities re-examined}
\vspace{0.2cm}

In this section we first discuss QED without Dirac monopoles.
For the massless Abelian photon gauge field $A_{\mu}$,
we make the following orthogonal expansion
$$
A_\mu={\cal A}_\mu + \tilde{\cal A}_\mu ,~~~~~
{\cal A}_\mu\equiv T_{\mu\nu}A^{\nu} ,~~~~~
\tilde{\cal A}_\mu\equiv L_{\mu\nu}A^{\nu} ,           
\eqno(1)
$$
where $T_{\mu\nu}$ and $L_{\mu\nu}$ are respectively the " transverse " and
" longitudinal " projection operators, which are
defined in coordinate space as
$~~~~
 T_{\mu\nu}=g_{\mu\nu}-\partial_\mu\partial_\nu /\partial^2,
{}~~~~L_{\mu\nu}= \partial_\mu\partial_\nu /\partial^2 ~~~~$,
or in momentum space as
$~~~~ T_{\mu\nu}=g_{\mu\nu}-k_\mu k_\nu/k^2,~~~~
L_{\mu\nu}=k_\mu k_\nu/k^2~~~~$. One should notice that the
operators $~T_{\mu\nu}~$ and $~L_{\mu\nu}~$ are non-local in
coordinate space but they give rise to {\it no} problem in
QED without monopoles, since in this case the gauge field is
regular everywhere.
Also it is well-known that the standard loop-corrections,
always calculated in momentum space, are non-local in
coordinate space. Further discussions on the case with monopoles
will be presented in Sec.2.

We emphasize that {\it  in any Lorentz frame
the $2$ physical transverse polarization states are always
included in the vector field ${\cal A}_{\mu}$ while the vector field
$ \tilde{\cal A}_{\mu}$ is just the unphysical
pure gauge field which has no
kinetic energy term and can be gauge-transformed away}.~
[After including Dirac monopoles, a detailed explanation on the gauge
transformations is given in the 2nd paragraph of Sec.4.1.]
The appearance of the unphysical pure gauge
field $ \tilde{\cal A}_{\mu}$ in the Lagrangian just makes the gauge
invariance manifest. Usually the coupling of both ${\cal A}_{\mu}$ and
$ \tilde{\cal A}_{\mu}$ to the matter field is denoted by the same physical
constant $e$. We notice that in fact the coupling of unphysical
$ \tilde{\cal A}_{\mu}$ can be any constant $\tilde{e}$ without affecting
the physical results of QED. In a covariant gauge, just the
 {\it gauge invariance itself } guarantees
that all $\tilde{e}$-dependent terms and also all $\xi$-dependent terms
 cancel one another completely and do
{\it not} contribute to any physical quantities such as the physical
$S$-matrix elements.

Consider first the QED Lagrangian without Dirac monopoles
$$
{\cal L}=-\frac{1}{4}F_{\mu\nu}F^{\mu\nu}+\bar{\psi} i \gamma_\mu
D^\mu\psi-m\bar{\psi}\psi                                     
\eqno(2)
$$
with
$$
\begin{array}{l}
D_\mu = (\partial_\mu-i e{\cal A}_\mu
-i{\tilde e}\tilde{\cal A}_\mu) , \\
F_{\mu\nu} = i e^{-1}[D_{\mu},D_{\nu}]
           = \partial_{\mu}{\cal A}_\nu-\partial_\nu{\cal A}_\mu ,
\end{array}
\eqno(3)
$$                                                            
where for clarity we have explicitly distinguished the unphysical
gauge coupling ${\tilde e}$ from the physical gauge coupling $e$.
Whether they are equal or not is irrelevant.
The above Lagrangian is then invariant under the following
$U(1)$ gauge transformations:
$$
\begin{array}{l}
\psi'(x)=U(\theta)\psi(x) ,~~~~~
\bar{\psi}'(x)=\bar{\psi}(x)U^{-1}(\theta) , \\
{\cal A}'_\mu(x)={\cal A}_\mu(x) , ~~~~~~~~
\tilde{\cal A}'_\mu(x)=\tilde{\cal
A}_\mu(x)-\tilde{e}^{-1}\partial_\mu\theta(x) ,                
\end{array}
\eqno(4)
$$
where $U(\theta)\equiv e^{-i\theta(x)}$. The transformations
for the gauge field can be combined together
$$
A'_\mu(x)=A_\mu(x)-\tilde{e}^{-1}\partial_\mu\theta(x) ~.       
\eqno(5)
$$
So we find that the coupling constant appearing in the above gauge
transformation of $A_{\mu}$ is actually the
{\it unphysical coupling $\tilde{e}$}. In fact this is not surprising
because the possibility of making gauge transformation for $A_{\mu}$
is just a reflection of its surplus unphysical degree of freedom
which is proportional to
$ \epsilon^{\mu}_L + \epsilon^{\mu}_S \propto k^{\mu} $,
where $ \epsilon^{\mu}_L $ and $ \epsilon^{\mu}_S $ are respectively
the unphysical longitudinal- and scalar-polarization vectors of
photon field.
It is clear that the ordinary gauge transformation is only a special
case of our generalized formulation with a special choice for the
unphysical coupling  $\tilde{e}$, i.e. $\tilde{e}=e$.
But the above formulation
describes the $~U(1)~$ gauge invariance more precisely
and reveals the essence
of the $~U(1)~$ local gauge invariance in a deeper way.
Also it is now clear that
ordinary gauge transformations with the special choice $\tilde{e}=e$ in
some sense {\it underdescribe} the $~U(1)~$ local gauge invariance.

We can write the corresponding formulation for
the non-integrable phase factor as
$$
P(x_2,x_1;C) = \exp\left[ i e\int^{x_2}_{x_1}{\cal
A}_\mu(x)dx^\mu+i\tilde{e}\int^{x_2}_{x_1}
\tilde{\cal A}_\mu(x)dx^\mu\right],
\eqno(6)                                                         
$$
where the line integral is along a path $C$ from $x_1$ to $x_2$
and we take $~{\hbar}=c=1~$.  For a closed loop C, (6) gives
$$
P(x,x;C)=\exp\left[ ie\int\int_S F_{\mu\nu}dS^{\mu\nu}\right] ~,
\eqno(7)
$$

\noindent
where $S$ is the surface bounded by the loop $C$ and it is clear that
(7) is $~U(1)~$ {\it gauge invariant} as expected and the unphysical
pure gauge field $~\tilde{\cal A}_{\mu}~$ cannot contribute to this
gauge-invariant physical quantity.
But this is valid only if both $~{\cal A}_{\mu}~$
and $~\tilde{\cal A}_{\mu}~$ are {\it not singular} on the surface $S$.
[ In the next section, we shall see that for Dirac monopoles, the
unphysical singularity (Dirac string) appears in both
$~{\cal A}_{\mu}~$ and $~\tilde{\cal A}_{\mu}~$ so that we cannot simply
use Stokes' theorem to obtain (7). In this case, the physical quantity
(7) is actually always ill-defined
due to the unphysical string singularity and
the arbitrariness of the unphysical gauge coupling $\tilde{e}$. This
will be examined in Sec.3. ]
In general, (6) gauge-transforms in the usual way
$$
P'(x_2, x_1;C)=U(\theta(x_2))P(x_2,x_1;C)U^{-1}(\theta(x_1)) ~.    
\eqno(8)
$$

Next, we re-examine the quantization process and will show that
{\it the renormalization
for the unphysical gauge coupling $\tilde{e}$ is completely different from
that of the physical gauge coupling $e$}. Consider the covariant Lorentz
gauge with the gauge fixing term
$$
{\cal L}_{gf} = -\frac{1}{2\xi}(\partial^{\mu}A_{\mu})^2
              = -\frac{1}{2\xi}(\partial^{\mu}
                 \tilde{{\cal A}}_{\mu})^2~~ ,
\eqno(9)                                                       
$$

\noindent
which clearly shows that the gauge fixing term is just a constraint on the
unphysical pure gauge field part $\tilde{{\cal A}}_{\mu}$ in $A_{\mu}$.
Then the generating functional of Green functions for QED
in the covariant gauge is
$$
\begin{array}{l}
Z[J,\tilde{J},I,\bar{I}] = \exp i W[J,\tilde{J},I,\bar{I}]\\ [0.5cm]
=\int {\cal D}\psi {\cal D}\bar{\psi}{\cal D}A_{\mu}
\exp i\int d^4x [{\cal L}+{\cal L}_{gf}
+J^{\mu}{\cal A}_{\mu} + \tilde{J}^{\mu}\tilde{\cal A}_{\mu}
+\bar{I}\psi + \bar{\psi}I] ~~,
\end{array}
\eqno(10)                                                    
$$

\noindent
where $~~{\cal L}~~$ is given by (2)(3).
[ Here we have noticed that in the path integral formalism
the true, independent dynamical functional integral variable
is still $A_{\mu}$ ( $={\cal A}_{\mu}+\tilde{{\cal A}}_{\mu}$ )
instead of the constrained fields
${\cal A}_{\mu}$ and $\tilde{{\cal A}}_{\mu}$,  respectively. ]
Then we define the classical fields as
$$
{\cal A}_{\mu}=\frac{\delta W}{\delta J^{\mu}}~,~~~
\tilde{\cal A}_{\mu}=\frac{\delta W}{\delta \tilde{J}^{\mu}}~,~~~
\psi = \frac{\delta W}{\delta\bar{I}}~,~~~
\bar{\psi} = -\frac{\delta W}{\delta I}~,
\eqno(11)                                                      
$$
where we do not distinguish the notations for classical fields and the
functional integral variables but we bear this difference in mind.
Making the Legendre transformation,
$$
\Gamma [{\cal A}_{\mu},\tilde{\cal A}_{\mu},\psi,\bar{\psi}]
=W[J,\tilde{J},I,\bar{I}]
- \int d^4x [J^{\mu}{\cal A}_{\mu} + \tilde{J}^{\mu}\tilde{\cal A}_{\mu}
+\bar{I}\psi + \bar{\psi}I] ~~,
\eqno(12)                                                       
$$
with
$$
\frac{\delta\Gamma}{\delta {\cal A}_{\mu}}=-J^{\mu} ~,~~~
\frac{\delta\Gamma}{\delta\tilde{\cal A}_{\mu}}=-\tilde{J}^{\mu} ~,~~~
\frac{\delta\Gamma}{\delta\psi}=\bar{I} ~,~~~
\frac{\delta\Gamma}{\delta\bar{\psi}}=-I ~.
\eqno(13)                                                     
$$
The functional $~Z~$ is
invariant under the gauge transformation (4), i.e.
$$
\int d^4x~[(\xi\tilde{e})^{-1}\partial^2\partial^{\mu}\tilde{\cal A}_{\mu}
-i\bar{I}\psi +i\bar{\psi}I + \tilde{e}^{-1}\partial^{\mu}\tilde{J}_{\mu}]
=0 ~,
\eqno(14)                                                     
$$
in which the fields are classical fields. Substituting (13) into (14)
we re-derive
the following generating equation for Ward-Takahashi (WT) identities
$$
(\xi\tilde{e})^{-1}\partial^2\partial_{\mu}\tilde{{\cal A}}^{\mu}
-\tilde{e}^{-1}\partial^{\mu}\frac{\delta\Gamma}
{\delta \tilde{{\cal A}}^{\mu}}
= i\left[\frac{\delta\Gamma}{\delta\psi}\psi +
\bar{\psi}\frac{\delta\Gamma}{\delta\bar{\psi}}\right] ~.
\eqno(15)
$$
By acting $~~\delta/\delta\tilde{\cal A}^{\nu}(y)~~$
and $~~\delta^2/[\delta\psi (z)\delta\bar{\psi}(y)]~~$
on (15) respectively and setting all external sources to zero,
the following two WT identities are derived after the
Fourier transformations:
$$
\begin{array}{l}
k^{\mu} i\tilde{D}^{-1}_{\mu\nu}(k) = -\xi^{-1}k^2k_{\nu} ~,\\
(p_{\mu}' - p_{\mu}) \Lambda^{\mu} (p', p)
 = i S^{-1}(p') - i S^{-1}(p) ~,
\end{array}
\eqno(16a,b)
$$
where
$$
i\tilde{D}^{-1}_{\mu\nu}(k) = \int_{FT}
\delta^2\Gamma /[\delta
{\tilde{\cal A}}^{\nu}(y) \delta {\tilde{\cal A}}^{\mu}(x)] ,~~~~
\tilde{e}\Lambda_{\mu}(p',p) \equiv \int_{FT} \delta^3\Gamma /
[\delta
\psi(z)\delta\bar{\psi}(y) \delta \tilde{{\cal A}}^{\mu}(x)] ,
\eqno(16c,d)                                              
$$
($\int_{FT}$ denotes the Fourier transformation)
and $S(p)$ is the full fermion propagator. Also we can easily find that
$$
\int_{FT} \delta^3\Gamma / [\delta
\psi(z)\delta\bar{\psi}(y) \delta {\cal A}^{\mu}(x)]
= e\Lambda_{\mu}(p',p)  ~.
\eqno(16e)                                                  
$$

To do the renormalization, we define
$$
\begin{array}{l}
\psi=Z_2^{\frac{1}{2}}\psi_R ~,~~~
\bar{\psi}=Z_2^{\frac{1}{2}}\bar{\psi}_R ~,~~~
{\cal A}^{\mu} = Z_3^{\frac{1}{2}}{\cal A}^{\mu}_R ~,~~~
\tilde{{\cal A}}^{\mu} = \tilde{Z}_3^{\frac{1}{2}}\tilde{{\cal
A}}^{\mu}_R ~,\\
m=Z_mm_R ~,~~~ e=Z_ee_R ~,~~~ \tilde{e}= Z_{\tilde{e}}\tilde{e}_R ~,~~~
 \xi = Z_{\xi}\xi_R ~,
\end{array}                                                  
\eqno(17)
$$
where we notice that the unphysical pure gauge field $\tilde{{\cal A}}_{\mu}$
have no physical asymptotic state and can never mix with the
transverse gauge field ${\cal A}_{\mu}$ since
$ ~~<0\mid T{\cal A}_{\mu}(x)\tilde{{\cal A}}_{\nu}(y)\mid 0> = 0~~$.
So, {\it $\tilde{{\cal A}}_{\mu}$ actually has an independent wavefunction
renormalization constant $\tilde{Z}_3$ }.

We rewrite (2) as
$$
{\cal L}=Z_3 \frac{-1}{4}F_{R\mu\nu}F_R^{\mu\nu} +
Z_2\bar{\psi}_R (i{\not\partial} - Z_mm_R){\psi}_R +
Z_1e_R{\cal A}_R^{\mu}\bar{\psi}_R\gamma_{\mu}\psi_R +
\tilde{Z}_1\tilde{e}_R\tilde{{\cal A}}_R^{\mu}\bar{\psi}_R\gamma_{\mu}\psi_R.
\eqno(18)
$$
Then we have
$$
Z_e = Z_1 Z_2^{-1} Z_3^{-\frac{1}{2}},~~~~~~
Z_{\tilde{e}} = \tilde{Z_1} Z_2^{-1} \tilde{Z_3}^{-\frac{1}{2}}.
\eqno(19)
$$
The WT identity (16a) {\it only requires} that after renormalization
$~~~~Z_{\xi} = \tilde{Z}_3 ~~~~$,
where {\it either $Z_{\xi}$ or $\tilde{Z}_3$ but not both
can be arbitrarily chosen}.
Since (16a) shows that $\tilde{D}_{\mu\nu}$ has no loop correction at all,
the most natural and simplest choice is
$$
Z_{\xi} = \tilde{Z}_3 = 1 ~.
\eqno(20)                                                             
$$
The conventional choice of
$~~~~ Z_{\xi} = \tilde{Z}_3 = Z_3 ~~~~$ is only another special choice.
The WT identity (16b)
and eqs.(16c,d,e) give $~~~Z_1 = \tilde{Z}_1 = Z_2~~~$.
So substituting this equation and (20) into (19) we get
$$
Z_e = Z_3^{-\frac{1}{2}} ,~~~~~~~
Z_{\tilde{e}} = Z_{\xi}^{-\frac{1}{2}} = 1 .
\eqno(21)
$$
Hence $Z_3$ can be determined,
for example, by the usual on-shell condition by requiring the
residue of the propagator for $~{\cal A}_R^{\mu}~$ to be unity at the
physical photon mass-pole $k^2 = 0$.
Thus we find that the physical coupling $e_R$ is
{\it running} as the renormalization
scale changes while the unphysical coupling $\tilde{e}_R$ is a
{\it renormalization-scale-independent} arbitrary constant
which always has a vanishing $\beta$-function.
 Hence, even if one chooses $~~\tilde{e} = e~~$
at tree level, the renormalization will make
$~~~e_R \neq \tilde{e}_R~~~$ up to
loop level since $\tilde{e}$ ($ \equiv \tilde{e}_R$) has no loop
correction and needs no renormalization whatsoever! In general we can have
$~~~Z_{\xi} = \tilde{Z}_3 = 1 + (arbitrary~~loop-order~~quantities)~~~$ and
this simply means that
$\tilde{e}$ has an arbitrary renormalized value different from
$e_R$. It is clear that no known fundamental physical principle can
insure that the unphysical coupling $\tilde{e}$ must be renormalized in
exactly the same way as the physical coupling $e$ and the usual choice of
$~~~ Z_{\xi} = \tilde{Z}_3 = Z_3~~~$ is only a special
one for convenience.

\newpage
\noindent
{\bf 3. Dirac quantization condition re-examined and the inconsistency
         of QED in the presence of Dirac monopoles}
\vspace{0.4cm}

\noindent
{\bf 3.1. Dirac quantization condition re-examined }

In the presence of Dirac monopoles, it is clear that the exact
$~U(1)_{em}~$ local gauge invariance must still be respected and thus our
precise formulation insures the arbitrariness of the non-observable
coupling $\tilde{e}$.

When Dirac monopoles are present, the quantization and
renormalization are much more complicated.
To the best of our knowledge, no consistent quantum
field theory for monopoles has been successfully constructed and generally
accepted as such.
The difficulties and doubts in consistently dealing with the analyticity,
Lorentz-invariance and crossing symmetry of
$S$-matrix were found by several authors about thirty
years ago$^{[6]}$. More recent attempts in this direction
have been reviewed by N. Craigie$^{[7]}$ who
discussed renormalization schemes different from that of Schwinger$^{[8]}$.
Here we wish to point out yet another basic difficulty
in the construction of monopoles's
quantum field theory. It is known that any quantum fluctuations
(i.e. loop-corrections)
which are always calculated in momentum space, are
however non-local in coordinate space (due to Heisenberg's uncertainty
principle) and will certainly not spare the singular region of $~A_{\mu}~$.
On the other hand, a singular
Dirac string in coordinate space will also be non-local in momentum space.
No-one knows how to deal with the quantum fluctuations of the Dirac string
nor to completely avoid these singularities at quantum-field-theory
level by using the well-known Wu-Yang formulation$^{[9]}$.

Due to above difficulties in the quantum field theory of monopoles,
here we keep
ourselves honest by only presenting our analysis at the level of quantum
mechanics. As in most previous considerations of this issue$^{[1][6][9][10]}$,
we believe that this restriction will not affect the generality of our
conclusion.
In this case, we can {\it explicitly} separate out the transverse
and longitudinal parts of monopole's gauge potential. Both are
{\it local} quantities and the general non-local operator $T_{\mu\nu}$ and
$L_{\mu\nu}$ are thus {\it unnecessary} here.

In this paper we only consider monopole potentials with Dirac string-type
singularities. Other kinds of singularities are studied
elsewhere$^{[5]}$. For clarity, our analysis is confined to the standard
Dirac formulation$^{[1,10]}$.
A corresponding examination in the Wu-Yang formulation$^{[9]}$
is left to a subsequent paper$^{[11]}$.

Let us consider a Dirac monopole $g$ with magnetic field
$~~ \vec{B}(x)=
\displaystyle\frac{g}{r^2}\frac{\vec{r}}{r} ~$,~
where $~ r = \mid\vec{x}\mid $~.
The magnetic field is related to the
monopole's gauge potential by
$~~\vec{B} = \vec{\nabla}\times \vec{A}~$ which implies that
$~A_{\mu}~$ cannot be regular everywhere and must contain some
singularities.  Since the physical $~\vec{B}~$ field is regular
everywhere except at the origin,
in the standard Dirac formulation$^{[10]}$ the above definition
is modified by adding the so-called Dirac string to cancel the
singularities in $~\vec{\nabla}\times \vec{A}~$ so that
the correct $~\vec{B}~$ field can be given. So one can write
$$
\begin{array}{l}
\vec{B} = \vec{\nabla}\times \vec{A}_{\ell}
          - \vec{b}(\ell,\vec{r})~,\\
\vec{b}(\ell,\vec{r}) = -4\pi g\int_{\ell}d\vec{x}~\delta^3(\vec{r}-\vec{x})~,
\end{array}
\eqno(22)                                     
$$
where $~\vec{b}(\ell,\vec{r})~$ denotes the contribution of the Dirac string
$\ell$ carrying a singular (return)
flux of strength $-4\pi g$
from infinity to the origin.
It is  shown$^{[10]}$ that the vector potential with singular line
$~\ell~$ can  then be expressed as
$$
\vec{A}(\vec{r})_{\ell} = -\int_{\ell}d\vec{x}\times\vec{B}(\vec{r}-\vec{x})
\eqno(23)                                           
$$
and two potentials with different singular lines are related by
the extended gauge transformation
$$
\begin{array}{l}
\vec{A}_{\ell'}(\vec{r}) = \vec{A}_{\ell}(\vec{r})
                 - \vec{\nabla}\Phi_{\ell',\ell}(\vec{r}),\\
\Phi_{\ell',\ell}(\vec{r}) = g\Omega(\ell,\ell';\vec{r})
\end{array}
\eqno(24)                                         
$$
where $~\Omega(\ell,\ell';\vec{r})~$ is the solid angle of the infinite
surface bounded by $\ell$ and $\ell'$, as seen from $\vec{r}$. Various
choices of the spanning surface will lead to values of $~\Omega~$ differing
by multiples of $4\pi$ but will yield the same $~\vec{\nabla}
\Omega~$, except on $\ell$ and $\ell'$ where $~\Omega~$  and
$~\vec{\nabla}\Omega~$ are always ill-defined.
The gauge
potential given above has a simple form when its singular line is chosen
to be a straight line. The simplest examples are the standard Dirac
solutions with singular lines along the negative and positive $z$-axes
respectively:
$$
(A_{\mp\hat{z}})_t=(A_{\mp\hat{z}})_{r}=(A_{\mp\hat{z}})_{\theta}=0~,
{}~~(A_{\mp\hat{z}})_{\varphi}
=\displaystyle\frac{g}{r}\frac{\pm 1-\cos\theta}{\sin\theta} ~.
\eqno(25)                                             
$$
Their transverse and longitudinal parts can be explicitly separated out as
$$
\begin{array}{l}
A^{\mu}_{\mp\hat{z}}={\cal A}^{\mu}_{\mp\hat{z}}+
                     \tilde{\cal A}^{\mu}_{\mp\hat{z}}~~,\\
{\cal A}^t_{\mp\hat{z}}=0~,  ~~~~
\vec{\cal A}_{\mp\hat{z}} =
-\displaystyle\frac{g}{r}\cot\theta\hat{\varphi}~~;\\
\tilde{\cal A}^t_{\mp\hat{z}}=0~,~~~~
\vec{\tilde{\cal A}}_{\mp\hat{z}}= \vec{\nabla}(\pm g\varphi).
\end{array}
\eqno(26)
$$                                                  
They are related by the gauge transformation
$$
\begin{array}{l}
{\cal A}^{\mu}_{\hat{z}}={\cal A}^{\mu}_{-\hat{z}}~,
\tilde{\cal A}^{\mu}_{\hat{z}}=\tilde{\cal A}^{\mu}_{-\hat{z}}
-\tilde{e}^{-1}\partial_{\mu}\alpha (x)~,\\
or,~~ A^{\mu}_{\hat{z}}=A^{\mu}_{-\hat{z}}
                        -\tilde{e}^{-1}\partial_{\mu}\alpha (x)~,\\
\alpha (x) = \tilde{e}g\Omega (-\hat{z},\hat{z};x), ~~~\Omega =2\varphi (x)~.
\end{array}
\eqno(27)                                                
$$
This is just a special case of the general relation (24).
Here we have noticed that
$~\vec{\nabla}\cdot\vec{\nabla}(\pm g\varphi )=0 ~$.
But since this gradient term can be
{\it arbitrarily changed} by $~U(1)_{em}~$
gauge transformations in order to
make Dirac strings pure gauge-artifact,
it must be unphysical and thus its coupling to
matter fields is still {\it non-observable.}
Actually this pure gradient term is
part of the unphysical $~U(1)_{em}~$ gauge degree of freedom. Otherwise (27)
is not a gauge transformation.
This is why it
should be attributed to the longitudinal field $~\tilde{\cal A}_{\mu}~$
associated with an unphysical coupling $~\tilde{e}~$. In principle, one
can call the above pure gradient term as "transverse" but its gauge coupling
is still non-observable and can be denoted as an arbitrary constant, say
$~e'~$, to avoid confusion. This will not affect our final conclusion.
Generally, when we consider other kind of
singular monopole potentials$^{[5]}$
we explicitly find that their pure gradient parts are {\it not}
divergenceless$^{[11,5]}$.
Corresponding to (27), the wavefunctin of an electrically charged particle
is gauge-transformed as $~~~\psi_{\hat{z}}(x) = e^{-i\alpha
(\varphi (x))}\psi_{-\hat{z}}(x)~~~$ in the regular region,
where the $~\varphi (x)~$ and $~~\varphi (x)+ 2\pi~~$
correspond to the same point $x$. Thus the single-valued property of
the wavefunctions requires $~~~ e^{-i\alpha (\varphi (x))} =
e^{-i\alpha (\varphi (x) + 2\pi)}~~~$.
Also the standard argument$^{[10]}$ shows
that there is a $~4\pi~$ discontinuity in the solid
angle $~\Omega(\ell,\ell';\vec{r})~$ so that the single-valued property of
the wavefunctions gives $~~e^{-i\tilde{e}4\pi g}=1~~$. So we must have
the following consistency condition
$$
\tilde{e} g=\frac{n}{2},~~~~(n=0,\pm 1, \pm 2,\cdots).
\eqno(28)                                           
$$
Here an important observation is that
{\it the physical magnetic charge $g$
is constrained by the non-observable gauge coupling $\tilde{e}$
for $n\neq 0$}.

One can further let the electric charge $e$ go around a closed loop
$C$ in the monopole's field. Then the wavefunctions $~\psi_{-\hat{z}}~$
and $~\psi_{\hat{z}}~$ get their phase factors given by
$$
\begin{array}{l}
\psi_{-\hat{z}}'(x) = \exp i[e\oint {\cal A}^{\mu}_{-\hat{z}}dx_{\mu}
+ \tilde{e}\oint \tilde{\cal A}^{\mu}_{-\hat{z}}dx_{\mu}]
\psi_{-\hat{z}}(x)~,\\
\psi_{\hat{z}}'(x) = \exp i[e\oint {\cal A}^{\mu}_{\hat{z}}dx_{\mu}
+ \tilde{e}\oint \tilde{\cal A}^{\mu}_{\hat{z}}dx_{\mu}]\psi_{\hat{z}}(x)~,\\
\psi_{\hat{z}}(x) = \exp [-i\alpha (x)]\psi_{-\hat{z}}(x) ~,~~~~
\psi_{\hat{z}}'(x) = \exp [-i\alpha (x)]\psi_{-\hat{z}}'(x) ~.
\end{array}
\eqno(29)                                               
$$
Since the DQC (28) already requires $~e^{-i\alpha (x)}~$ to be single-valued
and thus (29) gives another consistency condition
$$
\exp i\left[ e\oint {\cal A}^{\mu}_{-\hat{z}}dx_{\mu}
+ \tilde{e}\oint \tilde{\cal A}^{\mu}_{-\hat{z}}dx_{\mu}\right] =
{}~\exp i\left[ e\oint {\cal A}^{\mu}_{\hat{z}}dx_{\mu}
+ \tilde{e}\oint \tilde{\cal A}^{\mu}_{\hat{z}}dx_{\mu}\right] ~.
\eqno(30)                                                 
$$
Here we point out that (28) is the {\it pre-condition} of (30). Substituting
(26) into (30), choosing the closed loop $C$ in the horizontal plane with
$~\theta~$ and $~r\cos\theta~$ fixed, and then doing a trivial evaluation,
we get a consistency condition {\it identical to (28)}. ( Here one cannot
simply use the Stokes' theorem for the closed loop $C$ since
$~{\cal A}^{\mu}_{\mp\hat{z}}~$ and $~\tilde{\cal A}^{\mu}_{\mp\hat{z}}~$
are singular along the whole $z$-axis. )

At last we use the angular momentum approach$^{[10]}$
to {\it re-derive} DQC (28).
The equation of motion for electric charge $e$ in the monopole's field is
given by the Lorentz equation$^{[1][10]}$
$$
m\ddot{r}_{i} = e\dot{r}^{j}F_{ij}~,~~~ or,~~~
m\ddot{\vec{r}} = e\dot{\vec{r}}\times\vec{B}~,
\eqno(31)                                                
$$
where the coupling $e$ is the {\it physical} one associated with
tranverse gauge field since only the transverse field can contribute to
$~F_{\mu\nu}~$ (cf. (3)). From angular momentum conservation we get the
following expression of the total angular momentum in the
canonical formulation
$$
\begin{array}{l}
\vec{J} = \vec{r}\times m\dot{\vec{r}} - e g\hat{r}~,\\
m\dot{\vec{r}} = i\vec{D}~,
\end{array}
\eqno(32)                                                 
$$
where $~\vec{D}~$ is the covariant derivative defined in (3). Since the
theory is rotational invariant$^{[10]}$, for simplicity, consider
a monopole potential $\vec{A}_{-\hat{z}}~$
with its singular line along negative
$z$-axis (cf. (25)). Substituting $\vec{A}_{-\hat{z}}~$ into (32),
we get
$$
\begin{array}{l}
\vec{J} = \vec{r}\times\vec{p}
-g\left[ -\displaystyle\frac{\tilde{e}-e\cos\theta}{\sin\theta}
\hat{\theta} + e\hat{r}\right]~,\\
J_z = -i\frac{\partial}{\partial\varphi}-\tilde{e}g ~.
\end{array}
\eqno(33)                                                  
$$
When we quantize the theory, we expect the components of $~\vec{J}~$
to satisfy the usual angular momentum $~SU(2)~$ algebra which requires
the eigenvalues of $~J_i~$ to be {\it half integer}. Hence,
from (33) we again get a consistency condition {\it identical to
DQC (28)}.

\vspace{0.5cm}
\noindent
 {\bf 3.2. Inconsistency of Dirac monopoles}

In the standard Dirac formulation$^{[1][10]}$, we have already successfully
re-derived the DQC by three standard approaches
all leading to the same DQC (28).
We find that {\it only the non-observable gauge coupling $\tilde{e}$ can
appear in DQC}.
There are actually two kinds of solutions to DQC (28): $~~g=0~~$ with
$~~n=0~~$ and $g\neq 0$ with $~~n\neq 0~~$.  However the second
kind solution $~~g\neq 0~~$ implies the  following inconsistencies:
\begin{enumerate}
\item  
It constrains the physical magnetic charge $g$ with the non-observable
gauge coupling $\tilde{e}$, and is thus {\it conceptually inconsistent}
no matter which value of the unphysical $\tilde{e}$ one chooses.
(Indeed, no physical law can ever constrain a physical quantity with
a non-observable one.) We know of no way to rectify this situation.

\item  
Actually, this is a direct indication that Dirac strings cannot be
really made non-observable$^{[12]}$ since the DQC (28) originates from
the gauge transformations to arbitrarily move Dirac strings around
and the single-valuedness requirement of the electron wavefunctions.

\item  
Also, different choices for the unphysical
value of $\tilde{e}$ imply the physical $~g~$ must be arbitrarily varied.
This is physically unacceptable.

\item  
In the QED without monopole we have rigorously proved that the gauge coupling
$~\tilde{e}~$ is unphysical and arbitrary even up to quantum loop level and
this is a {\it direct} consequence of the exact $~U(1)_{em}~$ gauge invariance.
After Dirac monopoles are introduced in whichever way,
the exact $~U(1)_{em}~$ gauge invariance
must be respected. Inforcing any specific physical
value for an unphysical $~\tilde{e}~$
will violate the exact $~U(1)_{em}~$ gauge invariance and thus
implies that some new fundamental
principle should exist, which gives $\tilde{e}$ and the longitudinal
photon field physical meaning so that they become observable.
This is most unlikely.

\end{enumerate}
Hence we find that {\it the unique physically consistent solution}
to the ~DQC~ (28)~ can only be $~~~~ g = 0 ~~~$.
{\it This can be viewed as a theoretical evidence against
the existence of the Dirac monopoles}.

We further point out that the appearance of the DQC (28) is
essentially due to one's inability to define $A_{\mu}$ of Dirac
monopole globally without singularity.
To make the string-like singularity non-observable,
one {\it must make use of the gauge freedom}. So it is not surprising
that the unphysical coupling $~\tilde{e}~$ associated with longitudinal
gauge freedom naturally appears in the DQC (28) in all standard derivations.
This shows that the Dirac strings cannot be
really non-observable as desired.
It is clear that such a problem
does not arise for the 't Hooft-Polyakov monopole$^{[3]}$
which is a topological soliton with a finite core size determined
by the Higgs mass and has {\it no} singularity in its gauge potential
so that the mysterious "Dirac veto" is {\it unnecessary} there.
Moreover, in this case, one can easily check that the associated DQC emerges
{\it automatically } and involves only the physical tranverse $~U(1)_{em}~$
gauge coupling $e$.
The Aharonov-Bohm effect$^{[13]}$ is also free from such
a problem. An alternative, independent
proof of the above conclusions is presented in the next section,
which avoids entirely the introduction of
the non-local orthogonal expansion (1).

\vspace{0.5cm}
\noindent
{\bf 4. An alternative proof of the inconsistency}
\vspace{0.2cm}

In this section we present an alternative proof of the
inconsistency of QED in the presence of Dirac monopoles. We noticed that
the singular Dirac string$^{[1]}$ in the monopole gauge potential is purely
a gauge-artifact. It is just the gauge freedom which
allows us to arbitrarily move the string around without any physical effect,
provided that a consistent condition
--- Dirac quantization condition (DQC)$^{[1]}$ is
satisfied. By introducing another unphysical pure gauge field into QED, we
find it possible to attribute part of the singularities to this pure
gauge field and thus the corresponding DQC involves the unphysical
gauge coupling associated with this pure gauge field. So the physically
consistent solution to both the original DQC and this new DQC can only
be a vanishing magnetic charge. In Sec.4.1, a generalized QED Lagrangian with
an enlarged local gauge symmetry $~U(1)_{em}\times U(1)_{\Theta}~$ is proved
to be identical to standard QED up to the quantum-field-theory-level. Of
course,
the gauge coupling associated with this pure gauge field in the
$~U(1)_{\Theta}~$ group is shown to be entirely arbitrary. Two independent
DQCs are carefully derived in Sec.4.2
and some conclusions are  given in Sec.4.3.

\vspace{0.5cm}
\noindent
{\bf 4.1. A generalized QED Lagrangian and the Ward-Takahashi identities}

An Abelian or non-Abelian  global symmetry can always be localized by
introducing an {\it unphysical pure gauge field}, which has no kinetic term
and can be gauge-transformed away.
A dynamical gauge field is only a natural generalization
and at present its existence can be determined only by experiments.
A pure gauge field is {\it sufficient and necessary }
to insure the ordinary local gauge invariance.
This may be why without discovering
the corresponding dynamical gauge fields
we have observed a lot of global symmetries (such as
the lepton and baryon numbers conservations) which had been independently
tested at different local places.
Besides the electric charge conservation, standard QED
has an extra global $~U(1)~$ symmetry which is the electron number
conservation. In the following we shall localize this extra
global $~U(1)~$ symmetry by introducing a pure gauge field.
One should notice that only the physical gauge coupling
associated with a dynamical gauge field can be related to its global charge
and the unphysical gauge coupling associated with the pure gauge field has
nothing to do with the global charge since it is non-observable and the
corresponding pure gauge field can be completely  gauge-transformed away.

But when including monopoles, we should carefully distinguish
two essentially different situations.
In the Dirac monopole case, a singular gauge transformation must be allowed
in order to arbitrarily move the Dirac string and thus make it non-observable
as desired. This singular $~U(1)~$ gauge transformation
(which is usually called as an "extended" gauge transformation$^{[14]}$) can
thus arbitrarily  change the pure gradient part of the monopole's gauge field
or even entirely transform it away while leaving the physical magnetic field
invariant. This is in sharp contrast with the case of the spatially extended
't Hooft-Polyakov monopole$^{[3]}$ which, as finite energy solution to the
spontaneously broken gauge theories, is naturally singularity-free at
the beginning. All allowed regular gauge transformations cannot rid of
the pure gauge field (or even change their homotopy class). Furthermore
any singular gauge transformation which transforms the pure gauge field away
must be forbidden since it leaves a vanishing magnetic field.

In this section we first discuss QED without Dirac monopoles.
Consider the following generalized QED Lagrangian
$$
{\cal L}=-\frac{1}{4}F_{\mu\nu}F^{\mu\nu}+\bar{\psi} i \gamma_\mu
D^\mu\psi-m\bar{\psi}\psi                                     
\eqno(34)
$$
with
$$
\begin{array}{l}
D_\mu = (\partial_\mu-i e A_\mu
-i{\tilde e}\partial_\mu\Theta ) ~, \\
F_{\mu\nu} = \partial_{\mu}\bar{A}_{\nu}-\partial_{\nu}\bar{A}_{\mu}
           = \partial_{\mu}A_\nu-\partial_{\nu}A_{\mu} ~, \\
\bar{A}_{\mu}\equiv A_{\mu} + \partial_{\mu}\Theta ~,\\
\end{array}
\eqno(34a)
$$                                                    
where  $~\bar{A}_{\mu}~$ is only a notation
in which the coefficient of $~\partial_{\mu}\Theta~$ is arbitrary
but can always be chosen to be unity
since the {\it unphysical} $~\Theta~$ field
has no kinetic term and can be arbitrarily rescaled without any physical
effect.
The above QED Lagrangian has a larger
local symmetry $~U(1)_{em}\times U(1)_{\Theta}~$, i.e. it is invariant under
the following two kinds of independent gauge transformations:
\begin{description}
\item[(i).]      
The $~U(1)_{em}~$ gauge transformation
$$
\begin{array}{l}
\psi '(x)= e^{-i\alpha (x)}\psi (x) ~,~~~
\bar{\psi}'(x) = \bar{\psi}(x)e^{i\alpha (x)}~;~\\
A_{\mu}' = A_{\mu} - e^{-1}\partial_{\mu}\alpha (x) ~,\\
\Theta ' = \Theta ~.
\end{array}
\eqno(35)                                        
$$
\item[(ii).]     
The $~U(1)_{\Theta}~$ gauge transformation
$$
\begin{array}{l}
\psi '(x)= e^{-i\eta (x)}\psi (x) ~,~~~
\bar{\psi}' (x) = \bar{\psi}(x)e^{i\eta (x)} ~; \\
A_{\mu}' = A_{\mu} ~,\\
\Theta ' = \Theta - \tilde{e}^{-1}\eta (x) ~.
\end{array}
\eqno(36)                                        
$$
\end{description}
Eq.(36) clearly shows that the unphysical pure gauge field can be
completely gauge-transformed away and thus our generalized QED
Lagrangian simply reduces to the standard QED Lagrangian.
Actually the standard QED is in the {\it "unitary gauge"} of eq.(34,34a),
in which the pure gauge field $~\Theta~$ has been transformed away.
Here it is clear that the gauge coupling $~e~$ and $~\tilde{e}~$ belong
to the two direct product group
$~U(1)_{em}~$ and $~U(1)_{\Theta}~$ respectively,
and thus are {\it independent of each other}.

{}From (34,34a), the definition of the magnetic field is
$$
\vec{B} = \vec{\nabla}\times\vec{\bar{A}}
        = \vec{\nabla}\times\vec{A} ~.
\eqno(37)                                           
$$
The nonintegrable phase factor is now expressed as
$$
P(x_2, x_1; C) = \exp\left[ ie\int^{x_2}_{x_1} A_{\mu}dx^{\mu}
    + i\tilde{e}\int^{x_2}_{x_1}
      \partial_{\mu}{\Theta}dx^{\mu}\right] ~.
\eqno(38)                                             
$$
When doing quantization, we need two gauge-fixing terms for two gauge
groups $~U(1)_{em}~$ and $~U(1)_{\Theta}~$ respectively, i.e.
$$
{\cal L}_{gf} = -\frac{1}{2\xi_{A}}F_1(A)^2
         -\frac{1}{2\xi_{\Theta}}F_2(\Theta)^2 ~.
\eqno(39)                                              
$$
For example, the gauge-fixing functions $~F_1(A)~$ and $~F_2(\Theta )~$
can be chosen as
$$
F_1(A)=\partial^{\mu}A_{\mu} ~,~~~~ F_2(\Theta ) = \partial^2\Theta ~.
\eqno(40)                                             
$$
Now we derive some new Ward-Takahashi (WT) identities
for the $~U(1)_{\Theta}~$ gauge group. By introducing the external sources
$~~~J_{\mu}A^{\mu} +K\Theta + \bar{I}\psi + \bar{\psi}I~~~$
in the generating
functional for Green functions and doing a $~U(1)_{\Theta}~$ gauge
transformation, we can easily
re-derive the following generating equation
$$                                         
(\xi_{\Theta}\tilde{e})^{-1}\partial^4\Theta + \tilde{e}^{-1}
\frac{\delta\Gamma}{\delta\Theta} =
i\left[\frac{\delta\Gamma }{\delta\psi }\psi
+\bar{\psi}\frac{\delta\Gamma}{\delta\bar{\psi}}\right] ~.
\eqno(41)
$$
{}From (41) we get the following two WT identities
$$                                                  
\begin{array}{l}
i\tilde{D}^{-1}(k) = -\xi^{-1}_{\Theta}k^4 ~,\\
(p_{\mu}'-p_{\mu})\Lambda^{\mu}(p',p) = iS^{-1}(p')-iS^{-1}(p) ~,
\end{array}
\eqno(42a,b)
$$
where
$$                                                  
i\tilde{D}^{-1}(k) = \int_{FT}
\delta^2\Gamma /[\delta
\Theta (y) \delta \Theta (x)] ~,~~~
\tilde{e}(p_{\mu}'-p_{\mu})\Lambda_{\mu}(p',p)
\equiv \int_{FT} \delta^3\Gamma /
[\delta\psi(z)\delta\bar{\psi}(y) \delta \Theta (x)] ~,
\eqno(42c,d)
$$
($\int_{FT}$ denotes the Fourier transform)
and $S(p)$ is the full fermion propagator. Also we can easily find that
$$                                                
\int_{FT} \delta^3\Gamma / [\delta
\psi(z)\delta\bar{\psi}(y) \delta A^{\mu}(x)]
= e\Lambda_{\mu}(p',p)  .
\eqno(42e)
$$

To perform the renormalization, we define
$$                                                   
\begin{array}{l}
\psi=Z_2^{\frac{1}{2}}\psi_R,~~~
\bar{\psi}=Z_2^{\frac{1}{2}}\bar{\psi}_R,~~~
A^{\mu} = Z_A^{\frac{1}{2}}A^{\mu}_R,~~~
\Theta = Z_{\Theta}^{\frac{1}{2}}\Theta_R,\\
m=Z_mm_R,~~~ e=Z_ee_R,~~~ \tilde{e}= Z_{\tilde{e}}\tilde{e}_R,~~~
 \xi_A = Z_{\xi_A}\xi_{AR},~~~ \xi_{\Theta}=Z_{\xi_{\Theta}}\xi_{\Theta R}~.
\end{array}
\eqno(43)
$$
Here the non-observable gauge coupling $~\tilde{e}~$ of the
pure gauge field $~\Theta~$ has an independent renormalization
constant $~Z_{\tilde{e}}~$.

We rewrite (34,34a) as
$$                                               
{\cal L}=Z_A \frac{-1}{4}F_{R\mu\nu}F_R^{\mu\nu} +
Z_2\bar{\psi}_R (i{\not\partial} - Z_mm_R){\psi}_R +
Z_1e_R A_R^{\mu}\bar{\psi}_R\gamma_{\mu}\psi_R +
\tilde{Z}_1\tilde{e}_R\partial^{\mu}\Theta_R\bar{\psi}_R\gamma_{\mu}\psi_R~.
\eqno(44)
$$
Then we have
$$
Z_e = Z_1 Z_2^{-1} Z_A^{-\frac{1}{2}},~~~~~~
Z_{\tilde{e}} = \tilde{Z_1} Z_2^{-1} Z_{\Theta}^{-\frac{1}{2}}.
\eqno(45)                                                       
$$
The WT identity (42a) {\it only requires} that, after renormalization,
$~~~~Z_{\xi_{\Theta}} = Z_{\Theta} ~~~~$,
where {\it either $~Z_{\xi_{\Theta}}~$ or $~Z_{\Theta}~$ but not both
can be arbitrarily chosen}.
Since (42a) shows that $\tilde{D}_{\mu\nu}$ has no loop correction at all,
the most natural and simplest choice is
$$
Z_{\xi_{\Theta}} = Z_{\Theta} = 1 ~.
\eqno(46)                                                         
$$
In general, we can choose
$~~~Z_{\xi_{\Theta}}=Z_{\Theta}=1 + (~arbitrary~loop-order~quantities~)
{}~~~$.    The WT identity (42b)
and eqs.(42d,e)(43) give $~~~Z_1 = \tilde{Z}_1 = Z_2~~~$.
So substituting this equation and (46) into (45) we get
$$
Z_e = Z_A^{-\frac{1}{2}} ~,~~~~~~~
Z_{\tilde{e}} = Z_{\xi_{\Theta}}^{-\frac{1}{2}} = 1 ~.
\eqno(47)
$$
In consequence we prove that the renormalization for $\tilde{e}$ is actually
arbitrary and may need no renormalization whatsoever. This is not surprising
since for the product groups $~U(1)_{em}\times U(1)_{\Theta}~$, the gauge
coupling $~\tilde{e}~$ of $~U(1)_{\Theta}~$ has nothing to do with the
the physical coupling $~e~$ of $~U(1)_{em}~$.

Finally, we emphasize again
that our above generalized QED is {\it identical to} standard QED,
even up to loop-level. Clearly, the introduction of a pure gauge
field which can be gauge-transformed away can have no physical effects.

\vspace{0.5cm}
\noindent
{\bf 4.2. Dirac quantization condition re-examined }

Following Sec.3 we still work in the standard Dirac
formulation$^{[1,10]}$.
Let us consider a Dirac monopole $g$ with magnetic field
$~~~~ \vec{B}(x)=
\displaystyle\frac{g}{r^2}\frac{\vec{r}}{r} ~~~$,
where $ r = \mid\vec{x}\mid $.
The magnetic field is related to the
monopole's gauge potential by
$~~\vec{B} = \vec{\nabla}\times \vec{\bar{A}}~$ which implies that
$~\bar{A}_{\mu}~$ cannot be regular everywhere and must contain some
singularities.  Since the physical $~\vec{B}~$ field is regular
everywhere except at the origin,
in the standard Dirac formulation$^{[1,10]}$, the above definition
is modified by adding the so-called Dirac string to cancel the
singularities in $~\vec{\nabla}\times \vec{\bar{A}}~$, so that
the correct $~\vec{B}~$ field is obtained.
Following the same steps as in Sec.3, we obtain the two simplest Dirac
solutions for $~~\bar{A}_{\mu} (\equiv A_{\mu} +\partial_{\mu}\Theta )~~$
with singular lines along the negative and positive $z$-axes,
respectively:
$$                                                    
(\bar{A}_{\mp\hat{z}})_t=(\bar{A}_{\mp\hat{z}})_{r}
=(\bar{A}_{\mp\hat{z}})_{\theta}=0~,~~(\bar{A}_{\mp\hat{z}})_{\varphi}
=\displaystyle\frac{g}{r}\frac{\pm 1-\cos\theta}{\sin\theta} ~.
\eqno(48)
$$
They are connected by the gauge transformation
$$                                                   
\bar{A}^{\mu}_{\hat{z}}= \bar{A}^{\mu}_{-\hat{z}}
                           -\partial^{\mu}(2g\varphi )~.
\eqno(49)
$$
{}From (35) and (36), we see that this can be regarded as
a gauge transformation of $~U(1)_{em}~$ with
$$                                                  
\begin{array}{l}
A^{\mu}_{\hat{z}} = A^{\mu}_{-\hat{z}}- e^{-1}\partial^{\mu}\alpha (x)~,~~~~
\alpha = 2e g\varphi ~~\\
\vec{A}_{\hat{z}}
=\displaystyle\frac{g}{r}\displaystyle\frac{1-\cos\theta}{\sin\theta}
= \vec{A}_{-\hat{z}}-\displaystyle\frac{2g}{r\sin\theta}\hat{\varphi}~~,\\
\Theta_{\hat{z}} = \Theta_{-\hat{z}} =0 ~~;
\end{array}
\eqno(50)
$$
{\it or}, a gauge transformation of $~U(1)_{\Theta}~$ with
$$                                                 
\begin{array}{l}
\vec{A}_{\hat{z}}= \vec{A}_{-\hat{z}}
=\displaystyle\frac{-g\cos\theta}{r\sin\theta}\hat{\varphi} ~~,\\
\Theta_{\hat{z}}=\Theta_{-\hat{z}} - \tilde{e}^{-1}\eta (x)~,~~~~
\eta = 2\tilde{e}g\varphi~~,\\
\Theta_{\hat{z}}= -g\varphi = -\Theta_{-\hat{z}} ~~.
\end{array}
\eqno(51)
$$

Now we can repeat the three standard approaches given in Sec.3 to derive
the DQC by using the above two kinds of gauge potentials
and their transformations in (50) and (51), respectively.
Thus, from (50) we just obtain the ordinary DQC
$$                                                 
e g =\frac{n}{2} ~, ~~(n=0, \pm 1, \pm 2, \cdots )~~;
\eqno(52)
$$
while from (51) we get an {\it independent new DQC}
$$                                                 
\tilde{e} g =\frac{k}{2} ~,~~(k=0,\pm 1, \pm 2, \cdots )~~,
\eqno(53)
$$
which has a similar form to (52) but has a {\it completely
different physical meaning}.
Here an important observation is that in (53)
{\it the physical magnetic charge $g$
is constrained by the non-observable gauge coupling $\tilde{e}$
for $k\neq 0$}. This is not surprising
since the {\it singular Dirac string is a pure gauge artifact and thus
can be naturally attributed to an unphysical pure gaue field.}
It is easy to check that for the case of
the 't Hooft-Polyakov monopole$^{[3]}$
the DQC (53) cannot be derived even if one introduces an extra unphysical
$~U(1)~$ pure gauge field, since there is no singularity. Also
the original consistent condition (52) is unnecessary and the electric
charge is automatically quantized in the 't Hooft-Polyakov monopole case.

\vspace{0.5cm}
\noindent
 {\bf 4.3. Inconsistency of Dirac monopoles}

In (52) and (53) the gauge couplings $~e~$ and $~\tilde{e}~$ belong to
two direct product $~U(1)~$ groups respectively
and thus are {\it independent of each
other} as we pointed out before.
There are actually {\it two possible} solutions to the original
DQC (52): $~~g=0~~$ for
$~~n=0~~$ and $g\neq 0$ for $~~n\neq 0~~$.  However, in our new
DQC (53) the {\it only} physically consistent solution is
$~~~g=0~~~$ for $~k=0~$, which is also a possible solution to DQC (52).
The nonvanishing solution $~~g\neq 0~~$ in (53) constrains the
physical magnetic charge $~g~$ with unphysical coupling $~\tilde{e}~$
and thus can never be
consistent as already analyzed in previous Sec.3.2. Hence we conclude that
the unique physically reasonable solution to both (52) and (53) is
$~~g=0~~$, which implies that no Dirac monopoles exist in nature.
Thus this alternative, independent proof strengthens our conclusion in Sec.3
from a different point of view.
Other inconsistencies of Dirac monopoles are presented
elsewhere$^{[4,5]}$.

In summary, we have proposed in this paper
two independent, new approaches to re-examine the consistency of
Dirac point monopoles. We found that in order to make use of
the unphysical gauge freedom to make the singular Dirac strings non-observable,
the re-derived DQC inevitably constrains the physical magnetic charge $~g~$
with an unphysical gauge coupling $~\tilde{e}~$ if $~~g\neq 0~~$.
This clearly shows that the Dirac strings cannot be consistently made
non-observable.
Our above analyses thus provide further explanations for the
negative experimental results from Dirac monopole searches$^{[2]}$.

\null
\noindent
{\bf Acknowledgement }
H.J. H and Z. Q thank
Professors T.D. Lee, Y.P. Kuang, R. Barbieri, C. Itzykson,
J.C. Taylor, M. Peskin and A.S. Goldhaber for helpful discussions
on an early version of this work. H.J. H and C.H. T are
grateful to Professors Lay Nam Chang and Holger B. Nielsen for
helpful discussions and kind suggestions.
H.J. H thanks Professor A.S. Goldhaber
for continued discussions and  kind suggestions on this subject.

\null
\noindent
{\bf References }
\begin{enumerate}
\item                                 
P.A.M. Dirac: Proc. Roy. Soc. {\bf A133}(1931)60. An extension of this
idea to include relativity was given in his another paper: Phys.Rev.
{\bf 74}(1940)817. \\
For complete lists of related references, see:\\
S. Torres, W.P. Trower: {\it A Complete Magnetic Monopole Bibliography
1269-1986}, VPI $\&$ SU Report, VPI-EPP-86-8, 1986;\\
A.S. Goldhaber, W.P. Trower:
{\it Resource Letter MM-1: Magnetic Monopoles},
 Am.J. Phys. {\bf 58}(1990)429.

\item                                  
Particle Data Group: Phys.Rev. {\bf D50}(1994)1793.

\item                                   
G. 't Hooft:  Nucl. Phys. {\bf B79}(1974)276;\\
A.M. Polyakov: Sov. Phys. JETP Lett. {\bf 20}(1974)194.

\item                                   
H.J. He, C.H. Tze: VPI-IHEP-93-08, 09, 10.

\item                                   
H.J. He, C.H. Tze: {\it Stringless, Multi-valued Gauge Potentials and
the Inconsistency of Dirac Monopoles }, VPI-IHEP-93-11, ( submitted ).

\item                                    
D. Zwanziger: Phys.Rev. {\bf 137}(1965)B647; \\
S. Weinberg: ibid, {\bf 138}(1965)B988;\\
A.S. Goldhaber: ibid, {\bf 140}(1965)B1407.

\item                                    
N. Craigie: "{\it Monopoles and their quantum fields}" in
{\it Theory and Detection of Magnetic
Monopoles in Gauge Theories }, p.85, ed. N. Craigie, World Scientific
Publishing Co. Pet. Ltd., 1986.

\item                                      
J. Schwinger: Phys.Rev. {\bf 144}(1966)1087; {\bf 151}(1966)1048,1055.

\item                                       
T.T. Wu, C.N. Yang: Phys. Rev. {\bf D12}(1975)3845.

\item                                        
For general reviews, see for examples: B. Zumino: "{\it Recent developments
in the theory of magnetically charged particles}" in {\it Strong and Weak
Interactions---Present Problems},
({\it Proc. 1966 Int. School of Physics 'Ettore Majorana'} ), p.711,
ed. A. Zichichi, Academic Press, New York and London; \\
P. Goddard,  D.I. Olive:  Rep.Prog.Phys. {\bf 41}(1978)1357.

\item                                         
H.J. He, Z. Qiu, C.H. Tze: {\it Unphysical Gauge Coupling and the Inconsistency
of Dirac Monopoles }, VPI-IHEP-93-14. The equivalence proof for the Dirac and
Wu-Yang formulations has appeared long time ago. See for example:
R.A. Brandt, J.R. Primack: Phys.Rev.{\bf D15}(1977)1175 and Ref.[14].

\item                                          
The non-observability of Dirac strings has also been questioned before
from very different points of view.\\
For example, see:
H.J. Lipkin and M. Peshkin: Phys.Lett. {\bf B179}(1986)109.

\item                                           
Y. Aharonov, D. Bohm: Phys.Rev. {\bf 115}(1959)485;\\
R.G. Chambers:        Phys. Rev. Lett. {\bf 5}(1960)3.

\item                                           
P. Goddard,  D.I. Olive:  Rep.Prog.Phys. {\bf 41}(1978)1357.

\end{enumerate}

\end{sf}
\end{document}